\documentclass[
reprint,
superscriptaddress,
longbibliography,
preprintnumbers,
nofootinbib,
nobibnotes,
amsmath,
amssymb,
aps,
prl,
eprint,
]{revtex4-2}

\usepackage{graphicx}% Include figure files
\usepackage{subfigure}
\usepackage{siunitx}
\usepackage{dcolumn}% Align table columns on decimal point
\usepackage{bm}% bold math
\usepackage{xspace}
\usepackage{hyperref}% add hypertext capabilities
\hypersetup{
    colorlinks,
    linkcolor={red!50!black},
    citecolor={blue!90!black},
    urlcolor={blue!80!black}
}
\usepackage[utf8]{inputenc} % allow utf-8 input
\usepackage[T1]{fontenc}    % use 8-bit T1 fonts
\usepackage{hyperref}       % hyperlinks
\usepackage{booktabs}       % professional-quality tables
\usepackage{amsfonts}       % blackboard math symbols
\usepackage{nicefrac}       % compact symbols for 1/2, etc.
\usepackage{microtype}      % microtypography
\usepackage{xcolor}         % colors
\usepackage{amsmath}
\usepackage{graphicx}
\usepackage{subfigure}
\usepackage{booktabs}
\usepackage{comment}
\usepackage[capitalise]{cleveref}

\usepackage[printonlyused]{acronym}

\newcommand{\sklearn}{\texttt{scikit-learn}\xspace}

\newcommand{\pycbc}{\texttt{PyCBC}\xspace}
\newcommand{\gstlal}{\texttt{GstLAL}\xspace}
\newcommand{\spiir}{\texttt{SPIIR}\xspace}
\newcommand{\mbta}{\texttt{MBTA}\xspace}
\newcommand{\cwb}{\texttt{CWB}\xspace}

\newacro{CP}{conformal prediction}
\newacro{MCP}{Mondrian conformal prediction}
\newacro{CBC}{compact binary coalescence}
\newacro{FAR}{false alarm rate}
\newacro{IFAR}{inverse \ac{FAR}}
\newacro{MDC}{mock data challenge}
\newacro{ROC}{receiver operator curve}
\newacro{AUC}{area under curve}
\newacro{SNR}{signal-to-noise ratio}
\newacro{ML}{machine learning}
\newacro{MLP}{multi-layer perceptron}
\newacro{LR}{logistic regression}
\newacro{NN}{neural network}
\newacro{CNN}{convolutional neural network}

\newcommand{\pastro}{\ensuremath{p_{\rm astro}}\xspace}
\newcommand{\maxifar}{maximum-IFAR\xspace}

\begin{document}

\title{Enhancing gravitational-wave detection: a machine learning pipeline combination approach with robust uncertainty quantification}

\author{Gregory Ashton}
\email{gregory.ashton@rhul.ac.uk}
\affiliation{Department of Physics, Royal Holloway, University of London}
\author{Ann-Kristin Malz}
\affiliation{Department of Physics, Royal Holloway, University of London}
\author{Nicolo Colombo}
\affiliation{Department of Computer Science, Royal Holloway University of London}

\date{\today}

\begin{abstract}
Gravitational-wave data from advanced-era interferometric detectors consists of background Gaussian noise, frequent transient artefacts, and rare astrophysical signals. Multiple search algorithms exist to detect the signals from compact binary coalescences, but their varying performance complicates interpretation. We present a machine learning-driven approach that combines results from individual pipelines and utilises conformal prediction to provide robust, calibrated uncertainty quantification. Using simulations, we demonstrate improved detection efficiency and apply our model to GWTC-3, enhancing confidence in multi-pipeline detections, such as the sub-threshold binary neutron star candidate GW200311\_103121.
\end{abstract}

\maketitle

Gravitational-wave astronomy is progressing from initial detection to routine observation.
As of GWTC-4.0, the fourth gravitational-wave transient catalogue \citep{GWTC4}, the LIGO Scientific, Virgo, and KAGRA (LVK) collaborations have detected over 200 signals arising from \ac{CBC} sources.
These sources are discovered using highly developed search algorithms (pipelines) that measure significance by comparing a detection statistic for the candidate against an empirical background distribution.
Candidates are initially ranked by a frequentist \ac{FAR}, but the pipeline outputs are then convolved with an astrophysical model of the \ac{CBC} population to produce a 
Bayesian \pastro~\citep{Farr:2013yna, Dent:2021aaa, Andres:2021vew, Ray:2023nhx}.
The LVK routinely uses five pipelines to detect signals.
Four of these (\gstlal \citep{Messick:2016aqy, Sachdev:2019vvd, Tsukada:2023edh, Cannon:2020qnf, Ewing:2023qqe, Sakon:2022ibh}, \mbta \citep{Adams:2015ulm, Aubin:2020goo}, \pycbc \citep{Allen:2004gu, Canton:2014ena, Usman:2015kfa, nitz_2017, Davies:2020tsx}, and \spiir \citep{Luan:2011qx, Chu:2020pjv}) use parameterised models of \ac{CBC} sources, while \cwb \citep{Klimenko:2015ypf} uses a wavelet model with weaker assumptions about the source type.
Moreover, there are also external teams that run independent searches \citep[see, e.g.][]{Nitz:2021zwj, Mehta:2023zlk}.

To date, a straightforward approach has been taken to combining the results from multiple pipelines: taking the maximum \pastro or \acl{IFAR} (\acsu{IFAR}: i.e. 1/\ac{FAR}) across the set of contributing pipelines.
For example, in the GWTC, candidates with at least one pipeline with $\pastro > 0.5$ are considered significant signals (with an estimated contamination rate from non-astrophysical 
 sources of 10-15\%, see, e.g. \citet{GWTC3}).
This powerful and straightforward approach does not require processing and enables the simple combination of independent catalogues.
However, multiple estimates of a candidate's significance and properties by different algorithms also present an opportunity: correlations between pipelines could be exploited to improve the overall detection efficiency beyond the current maximum approach.
This idea has already been explored for combining the pipelines to produce a unified \pastro \citep{Banagiri:2023ztt}, but this relies on accurate models of the signal and noise distributions.
In this work, we explore a new approach that combines pipelines using simple \ac{ML} models trained on labelled data.
However, the predictions from such models are uncalibrated and lack a quantified uncertainty.
Therefore, we augment our ML-combination pipeline by applying \ac{CP} \citep{vovk2005algorithmic, angelopoulos2021gentle} to provide quantified uncertainty measurements using labelled calibration data.
This approach is fast, computationally efficient, and requires only the simulation of the expected signal and noise distributions.
Our approach offers the capacity to learn the strengths and weaknesses of multiple pipelines without strict requirements on the underlying data products.
Thus, it can also be used to assess the performance of new pipelines or modifications to existing pipelines.
We restrict ourselves to the binary classification problem, signal or noise, but the work can be generalised to multi-class classification straightforwardly.

We use two standard \ac{ML} classification models: \ac{LR} and a \ac{MLP}; both are discussed in detail in the Appendix.
Each takes as input a feature vector $\vec{X}$ and returns a normalised set of probabilities $P$ for each label.
To train the models, we utilise the results from the recent \ac{MDC} study in advance of the LVK fourth observing run \citep{Chaudhary:2023vec}, where the \gstlal, \pycbc, \spiir, \mbta, and \cwb search pipelines were applied to a real-time data replay with added simulated signals.
The 40 days of data are taken from the LIGO Livingston and Hanford detectors \cite{LIGO} and the Virgo detector \cite{Virgo} during the third observing run (the KAGRA detector \cite{KAGRA:2018plz} was not in operation at this time).
From this \ac{MDC}, we take all candidates, excluding early-warning candidates, that the search pipelines upload, cluster in time (grouping all events within a \qty{1}{\s} window), and then filter all but the maximum \ac{SNR} candidate per pipeline.

This produces our feature data $\{\vec{X}_n\}$ where each row contains the per-pipeline features (including detection quantities such as the \ac{IFAR}, \ac{SNR} alongside estimates of the source properties such as the mass and spin; see the Appendix for details).
However, we do not include \pastro as a feature because the enhanced signal rate used in constructing the data means the pipeline \pastro values are not well calibrated.
For elements of the feature data where any given pipeline does not find a candidate with \ac{IFAR} > 1~hr, we enter zeros to fill in the missing data.

We then compare the candidate list with the times of known simulated signals and astrophysical signals known to be in the data and produce a ground-truth label set $\{Y_n\}$.
We find 9946 rows of data, with 5908 corresponding to simulated or real signals.
The number of signals in this data is significantly greater than the rate of detections expected for advanced-era detectors, as simulated signals were added to the data at a rate much higher than the anticipated astrophysical rate to stress-test the low-latency infrastructure.
With the data in hand, we then split the data into three subsets: 10\% for \ac{CP} calibration, 10\% for testing, and the remainder for training.

\begin{figure}
    \centering
    \includegraphics[width=\linewidth]{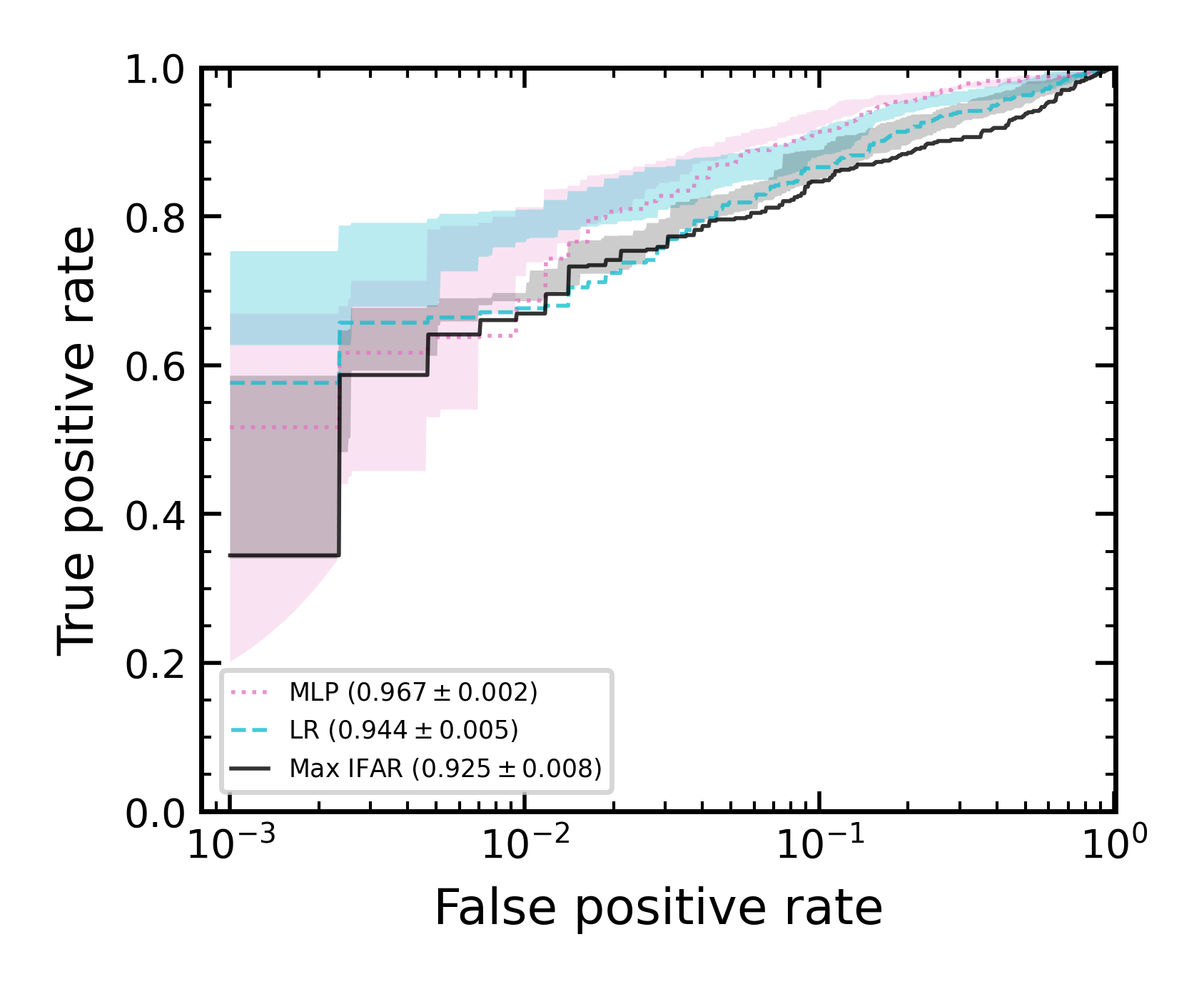}
    \caption{The \ac{ROC} for the \ac{LR} and \ac{MLP} ML-driven pipeline combination approaches applied to the test data; we also include as a comparison the standard \maxifar pipeline combination approach.
    For this test, we use all four pipelines contributing to the MDC and all features in our test data (see the Appendix for details).
    To investigate the uncertainty inherent in the \ac{ROC} curve, we run the study under different permutations of the training and test data.
    The solid lines indicate the \ac{ROC} calculated for a single permutation of the test data, while the shaded band marks the 90\% interval over the permutations.
    }
    We quantify the difference between combination approaches in the legend by providing the \ac{AUC} along with an estimate calculated under several training and test data permutations.
    \label{fig:roc}
\end{figure}

In \cref{fig:roc}, we compare the \ac{ROC} for the two ML pipeline combination approaches to the standard \maxifar method.
This demonstrates the potential of ML to improve the detection efficiency, as quantified by the \ac{AUC} provided in the legend.
Specifically, both ML approaches deliver an increase in the AUC above the level of uncertainties in the AUC as measured over the test data.
However, we note that comparing the uncertainty in the \ac{ROC} curve itself, the distributions do overlap, but their uncertainty envelopes are visually separated. 

Comparing the two ML approaches, \cref{fig:roc} demonstrates that the \ac{MLP} approach outperforms the simpler \ac{LR} model as measured by the AUC.
This is expected since the \ac{MLP} is more expressive: it can capture more complicated patterns due to the more involved underlying architecture.
However, while the \ac{LR} model is simpler, the results are easily interpreted. A simple inspection of the fitted coefficients can provide insight into the importance of individual features for each pipeline (see Appendix).
For the advantage of interpretability and a modest reduction in performance, we present only the results for the \ac{LR} model hereafter.

So far, we have demonstrated that an ML-driven pipeline combination approach can outperform a naive maximum-IFAR combination as measured by the \ac{ROC}.
However, in contrast to the results from combining individual search pipeline results, an ML pipeline combination does not provide a well-calibrated measure of the uncertainty.
This is important because a central aim for any method seeking to identify signals is to assess the significance of individual candidates.
As argued in \citet{Gebhard:2019ldz}, discussed in the context of using a \ac{CNN} as a search algorithm, the output of any \ac{ML} classifier is a function of the test data set and therefore is not necessarily calibrated to reality.
They conclude that ``\ac{CNN}s alone cannot be used to properly claim gravitational-wave detections''.

This difficulty is not unique to gravitational-wave astronomy; uncertainty quantification is a topic of interest in many high-stakes applications of \ac{ML} where predictions must be robust.
One approach to providing robust, well-calibrated predictions is \ac{CP}, a distribution-free approach that requires only exchangeability of the data and can be applied to any point predictor to produce statistically rigorous prediction regions \citep{vovk2005algorithmic, angelopoulos2021gentle}.
In the Appendix, we provide a brief introduction to \ac{CP}, but we have previously applied \ac{CP} to the problem of gravitational-wave astronomy \cite{Ashton:2024wae}, demonstrating how to calibrate individual pipelines.
We now extend that work to quantify uncertainty for an \ac{ML} combination pipeline.
Specifically, we apply standard label-conditional prediction using the complement of the \ac{LR} prediction probability as a non-conformity score (in \citet{Malz:2024zjd}, we explore alternative scores but do not find compelling advantages to use these in this work).

To measure the significance of an individual event within the \ac{CP} framework, we can use the confidence \cite{Shafer:2007}. In \citet{Ashton:2024wae}, we explored three possible definitions of the confidence, each with its own merits.
In this work, we will apply the ``Conditional confidence: signal'' which is defined as the minimum value of $\alpha$ such that the signal label is included in the prediction set $\Gamma^{\alpha}$.
We choose this conditional confidence because $i)$ unlike the standard definition of the confidence \cite{Shafer:2007}, it can be measured for any label on any test data, $ii)$ it can be generalised to the multi-class case trivially, and $iii)$ it enables the straightforward definition of a catalogue with a calculable purity by placing a threshold on the conditional confidence.

\begin{figure}
    \centering
    \includegraphics[width=0.49\textwidth]{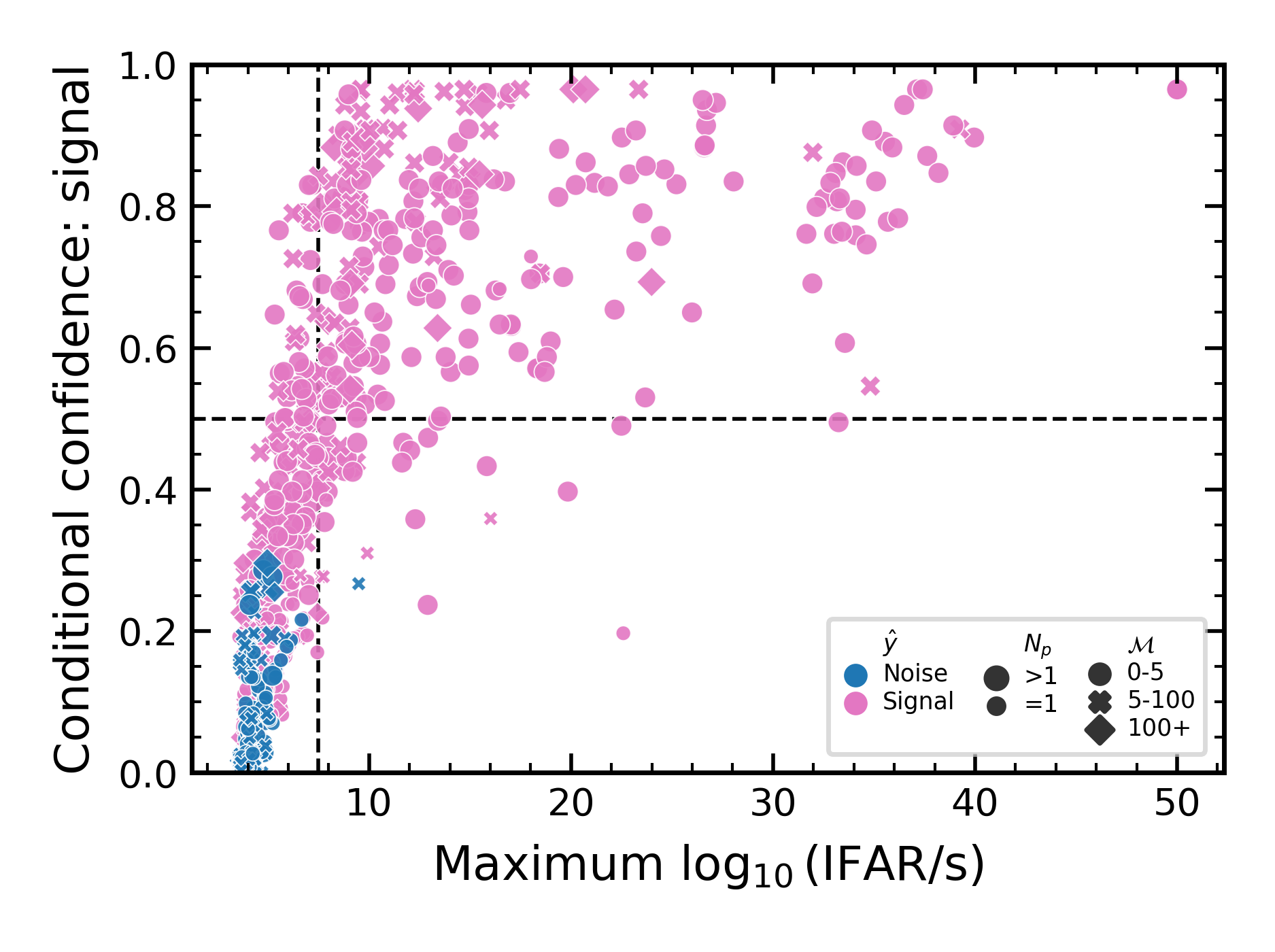}
    \caption{
    The conditional confidence in the signal label as measured by the \ac{LR} model and applied to the test data compared to the maximum \ac{IFAR}.
    We highlight the true label ($\hat{y}$) by the colour, the number of contributing pipelines ($N_p$) by the size, and the chirp mass ($\mathcal{M}$) range inferred by the highest-SNR pipeline by the symbol.
    A vertical dashed line marks a FAR threshold of 1 per year.
    }
    \label{fig:confidence}
\end{figure}

In \cref{fig:confidence}, we plot the conditional confidence against the maximum-IFAR for all test data points using the \ac{LR} model and then highlight the true label, number of contributing pipelines, and measured chirp mass \citep{LIGOScientific:2025hdt} of the signal.
Comparing the confidence with the maximum \ac{IFAR}, we observe that events identified by multiple pipelines tend to have higher confidence (events found by a single pipeline are mostly located at the lower edge of the distribution for a given IFAR).
Examining specific events, we observe a single false positive (using a standard threshold of 1 per year), which is assigned an IFAR of approximately $ 10^{10}$~s by the \gstlal pipeline.
Under the maximum-IFAR approach, such an event would be considered significant; however, the confidence of the event is found to be small ($\approx 0.35$) relative to other candidates found by multiple pipelines at a similar IFAR.
However, on the other hand, nearby this candidate, there is a low-mass event found by \gstlal and \pycbc, which is ranked with a similar confidence despite being found by multiple pipelines.
This suggests more work is needed to understand how the confidence is assigned and optimise it to better separate signals and noise.

A core assumption of \ac{CP} is that the calibration data and the test data are 
\emph{exchangeable} \citep{Shafer:2007}: given a collection of $N$ data points, the $N\!$ different orderings are equally likely.
For the problem of reasonably well-calibrated search pipelines studying a stream of data (e.g. from a given observing run), it seems reasonable that their results will be exchangeable.
I.e., we do not expect the meaning of the FAR and the measured parameters, such as mass, to vary throughout an observing run.
However, this may not be true if there are changes to the search pipeline or the instruments (say, we utilise examples from a previous observing run).
Therefore, care should be taken whenever the calibration data and test data are sourced differently (which will always be the case when studying real data, as we do not know the ground truth about the sources impacting our detectors). 
Moreover, careful investigation is needed to understand the importance of the relative numbers of signals and noise candidates in the calibration data.
For the demonstrations above, we have guaranteed exchangeability by randomly splitting the MDC data set.
We will now go beyond this test data set to study results from real searches for signals.

We study the candidate lists from the O3a and O3b observing runs published as part of the GWTC-3 catalogue \cite{gwt2p1_zenodo, gwtc3_zenodo}.
Specifically, this includes a list of the search pipeline output from the \cwb, \pycbc, \gstlal, and \mbta pipelines.
However, for \pycbc, a second search was performed targeting only BBH candidates; we excluded these results to improve the exchangeability of the training and test data.
Furthermore, we utilise only the measured \ac{IFAR}, \ac{SNR}, and chirp mass (except for \cwb); this is done to best ensure exchangeability, as the \ac{FAR} is calibrated and checked during pipeline development.
Nevertheless, we acknowledge that the pipelines were developed between O3 analyses and the MDC, so there are likely differences in their behaviour.
We take MDC data using this restricted feature set for use in training the \ac{LR} model and for calibration.

\begin{figure}
    \centering
    \includegraphics{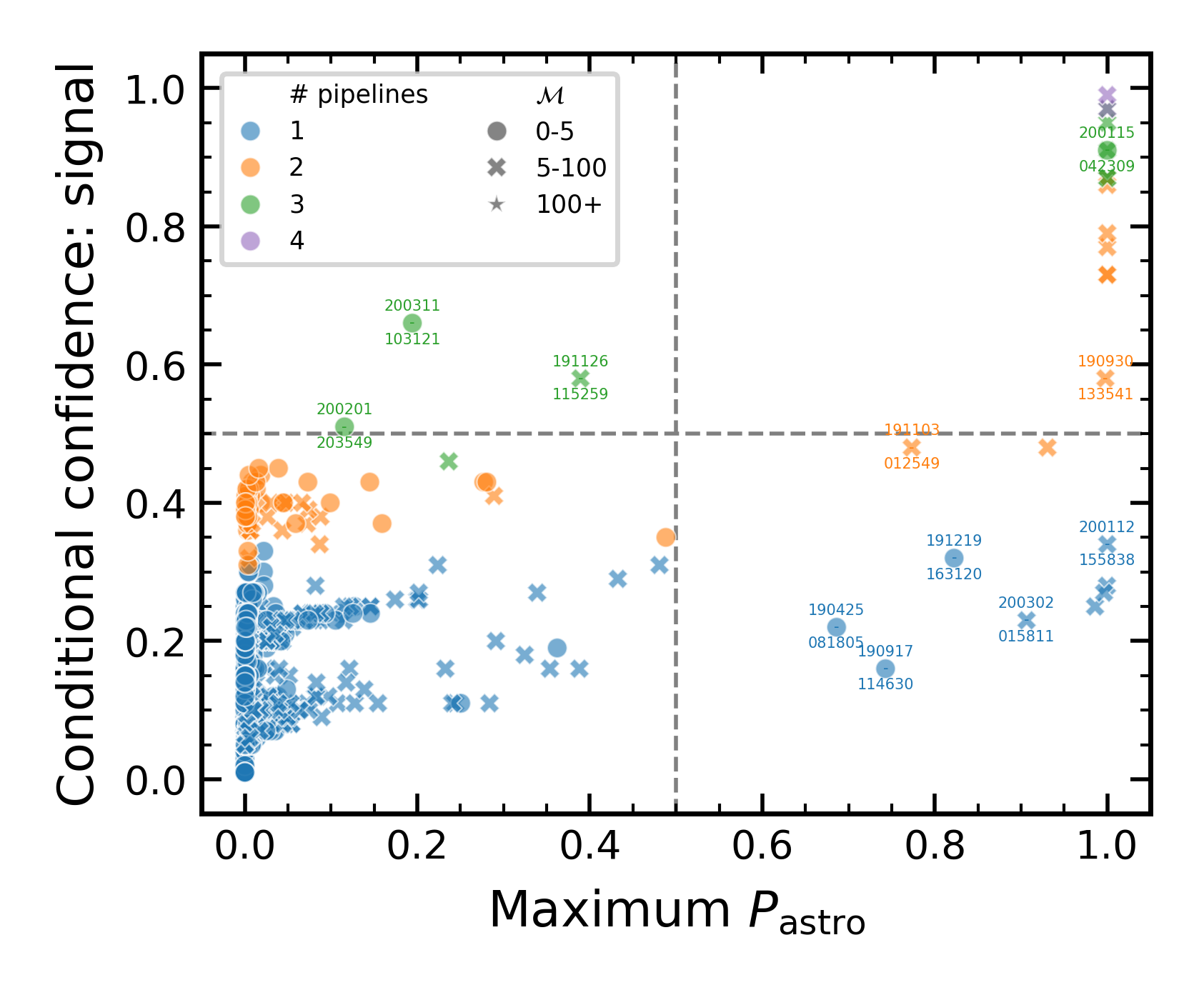}
    \caption{A comparison of the \pastro and conditional confidence using the \ac{LR} model trained on a subset of the MDC data.
    }
    \label{fig:gwtc3}
\end{figure}

In \cref{fig:gwtc3}, we plot the conditional confidence obtained from our \ac{LR} combination model against the maximum \pastro across pipelines.
We compare against \pastro here as this is the primary metric used in the GWTC to threshold for further analysis.
However, we note that \pastro is not included in the features used to train our \ac{LR} model.
This is because, in addition to the issues with the \pastro values within the MDC \citep{Chaudhary:2023vec}, while it is possible to use \pastro as a feature, this is one of the features we know can be non-exchangeable since the astrophysical population improves as we see more events.
Therefore, the population model used to calculate \pastro for the training data is different to that used to calculate \pastro for the test data.
As a result, the \pastro presented in \cref{fig:gwtc3}, contains information about the astrophysical population not available in the measurement of the conditional confidence.

From \cref{fig:gwtc3}, the four quadrants reveal an insight into the comparative performance of the traditional \pastro method and the \ac{CP} confidence.
First, we note that they are correlated: we have most of the data points in the top-right and bottom-left.
In the top-right quadrant, we see a cluster of events with a $\pastro \sim 1$ and confidence $\sim 1$ (see the Appendix).
Just below this cluster. we also find GW200115\_042309, one of the first detected NSBH signals \cite{NSBH} with a confidence of $\sim 0.9$, which was not detected by \cwb. Finally, we also find GW200209\_085452, a BBH candidate not found by \cwb.
In the bottom-left quadrant, we find sub-threshold candidates from both methods; we observe some stratification in the confidence, which is not yet understood.

In the bottom-right quadrant, we find candidates with $\pastro>0.5$ but confidence below 0.5.
Except in two cases with a confidence level of $\sim0.5$, these candidates are identified by only a single pipeline. For example, GW200302\_015811 was found by \gstlal in data from Hanford and Virgo, but the Virgo data had an SNR less than 4. 
Meanwhile, GW200220\_061928 is a high-mass candidate found only by \pycbc.
The candidate with the lowest confidence but highest \pastro is GW190917\_114630, found only by \gstlal in GWTC-2.1 with a \pastro of $\sim0.7$ \cite{GWTC2-1, GWTC3}.
Based on the source properties, this is most likely an NSBH \cite{GWTC2-1, GWTC3Pop}. However, its properties are also found to be inconsistent with the isolated binary evolution pathway \cite{Broekgaarden:2021hlu}.
Nearby this event, we also find GW190425\_081805, the second observed BNS \cite{GW190425}, which is similarly only found by the \gstlal pipeline (again, we report the updated \pastro from GWTC-2.1 \cite{GWTC2-1}.

Finally, we focus on the upper-left-hand quadrant: candidates above a confidence threshold of 0.5 but below a \pastro of 0.5, where we find three candidates.
First, GW191126\_115259 is a BBH candidate found by \gstlal, \pycbc, and \mbta with a maximum \pastro of 0.39 (in \pycbc), but a confidence of $\sim 0.6$.
Notably, this event is detected by the \pycbc-BBH search with a \pastro of 0.7 (these results are excluded from our test data set for reasons stated above).
Next, GW200311\_103121 and GW200201\_203549 both appear in the marginal candidate table of GWTC-3, and from a multi-component \pastro analysis, they are indicated (if real) to be a BNS and NSBH, respectively.
These events are given greater confidence relative to their maximum \pastro, which arises from the fact that three pipelines find them.
While we do not claim that our new metric is robustly tested enough to claim these as up-ranked detections, they demonstrate the increased significance possible from combining pipelines.
A re-analysis of this data, with better control over any systematic differences in pipeline behaviour, may yield the ability to trust the increased significance.
If confirmed, GW200311\_103121 would be only the third BNS signal detected, underscoring the importance of using all available information to assess significance.

In summary, we have introduced a new approach to pipeline combination, offering improvements over the current method that takes a simple maximum \pastro or \ac{IFAR}.
In this approach, we utilise \ac{ML} to learn the optimal combination from a set of training data in which the ground truth is known.
Utilising a recent \ac{MDC} \citep{Chaudhary:2023vec}, we demonstrated that simple off-the-shelf \ac{LR} or \ac{MLP} models can outperform the maximum-IFAR combination at the population level.
However, such an approach is limited by itself, as it lacks a robust measurement of prediction uncertainty for individual events.
Therefore, we introduce \ac{CP}, which can provide robust uncertainty measurements through an additional calibration data set.
We demonstrate the application of the \ac{CP} confidence to observed data from O3 and find a handful of events (most notably GW200311\_103121, a possible BNS event), where the combination approach yields increased confidence relative to \pastro, which stems from the multiple pipelines that identified the candidate.

For experts within the field, utilising the outputs from multiple pipelines is a standard process when assessing the significance of a candidate.
The combination approach proposed here does not replace that expertise but aims to enhance it, providing a single rigorous quantified uncertainty.
The key beneficiary of this approach is astronomers who use gravitational-wave alerts and the GWTC (e.g. to trigger observations or perform further studies).
A single statement of confidence in a candidate, which combines the parameter-space-dependent sensitivity of all pipelines, will provide clarity and interpretability.
We envision that the field could utilise this approach to combine search pipelines when producing a catalogue of events or low-latency alerts.
A single, easy-to-understand assessment of the multi-pipeline results will ease interpretation and potentially improve sensitivity.

\textbf{Acknowledgements}
We thank Sharan Banagiri, Francesco Salemi, Tito Dal Canton, and Surabhi Sachdev for discussions that helped in the development of this work and the two anonymous referees who provided useful feedback during the review.
We also acknowledge unpublished work by Nikolas Moustakidis, Theofilos Moustakidis, Erik Katsavounidis, and Deep Chatterjee that took a similar approach to our ML pipeline combination method.
Implementation of our ML models was done using \sklearn \cite{Buitinck:2013fcp}, while we utilise \texttt{NumPy} \cite{harris_2020}, \texttt{Pandas} \cite{reback2020pandas}, and \texttt{Matplotlib} \cite{Hunter:2007} for data handling and visualisation.
We also thank Michael Coughlin,
Deep Chatterjee, Tito Dal Canton, Reed Essick, Shaon
Ghosh, Sushant Sharma-Chaudhary, Max Trevor, and
Andrew Toivonen for the development of the MDC results used in this work. This material is based upon
work supported by NSF’s LIGO Laboratory, which is a
major facility fully funded by the National Science Foundation. 
The authors are grateful for computational resources provided by the LIGO Laboratory and supported by National Science Foundation Grants PHY-0757058 and PHY-0823459.

\bibliographystyle{apsrev4-2}
\bibliography{bibliography}

\onecolumngrid
\newpage

\thispagestyle{empty}
\begin{center}
{\large \textbf{Supplemental Material to “Enhancing gravitational-wave detection: a machine learning
pipeline combination approach with robust uncertainty quantification”}}
\end{center}

In this Supplemental Material, we provide additional details on several topics discussed within the Letter ``
Enhancing gravitational-wave detection: a machine learning pipeline combination approach with robust uncertainty quantification''.

\subsection{Machine-learning models}
In our \ac{ML} pipeline combination approach, we explore two different supervised \ac{ML} models and compare their performance: \ac{LR} and a \ac{MLP}.
In both cases, the \ac{ML} model takes as input a feature vector $\vec{X}$ and returns a normalised vector of probabilities $P = P(\vec{X}; \lambda)$, where $\lambda$ represents the free parameters of the respective model and the size of $P$ is equal to the number of classes (in the binary case, two).
\ac{LR} is a simple \ac{ML} model in which the input feature vector is converted to a probability by the sigmoid function $P(\vec{X}; \lambda) = \sigma(z)$ where $\sigma(z)$ is the sigmoid function and $z=\vec{w}\cdot\vec{X}+b$, combining the input feature vector with a vector of weights $w$ and a bias $b$.
Meanwhile, \ac{MLP} is an \ac{NN}, where layers of neurons transform the input feature vector to an output probability for each prediction class.
We use one hidden layer of 100 neurons and an ReLU (rectified linear unit) \cite{Agarap:2018uiz} activation function to calculate the output of each neuron.
Then, an output layer converts and combines the outputs from the hidden layer to probabilities using the sigmoid function.
Thus, the output from our \ac{MLP} model is again $P(\vec{X};\lambda)=\sigma(z)$ but now
$
    z=b^o+\sum_{k=1}^K w^o_k \, \mathrm{ReLU}\left(\vec{w}^{h}_k \cdot \vec{X}+b^h_k\right)\,,
$
where the weights and biases $\vec{w}^h$, $b^h$ correspond to the $K$ neurons in the hidden layer, while $w^o$, $b^o$ are associated with the output neuron.
We also explored several other \ac{ML} approaches in development but found only subtle differences. Therefore, we present two case studies that demonstrate these general trends.

Both models can be trained by defining an inverse loss function; we use the log-likelihood, calculated as
\begin{equation}
     \sum_{n=1}^N \left( Y_n\log(P(\vec{X}_n;\lambda))+(1-Y_n)\log(1-P(\vec{X}_n;\lambda))\right),
    \label{eq:log-likelihood}
\end{equation}
where $Y_n\in{0,1}$ are the known classifications (one for signal, zero for noise), $N$ is the size of the training dataset, the subscript $n$ labels the sample from the training dataset, and $P(\vec{X}_n;\lambda)$ is the output from the respective \ac{ML} model. 

During training, we use \texttt{L-BFGS-B} \citep{zhu1997algorithm}, a quasi-Newton code for bound-constrained optimisation, and \texttt{Adam} \citep{Kingma:2014vow}, a stochastic gradient descent optimiser, for \ac{LR} and \ac{MLP} respectively, to maximise the log-likelihood in equation \cref{eq:log-likelihood} and obtain $\hat{\lambda}$, the best-fit parameters.
Once the models have been trained, classification probabilities for new test data $\vec{X}'$ can be calculated as $P(\vec{X}'; \hat{\lambda})$.

\subsection{Per-pipeline features}
\label{sec:features}
The data available from the \ac{MDC} \citep{Chaudhary:2023vec} includes quantities summarising the significance of the candidate alongside estimates of the source properties.
For the template-based pipelines, we record the \ac{FAR}, \ac{SNR}, and $\chi^2$ (a discriminator used to separate astrophysical signals from transient noise based on the time-frequency behaviour, see \citet{Allen:2004gu} for the development and references to individual pipelines above for the current implementation).
Templated pipelines also provide source-parameter estimates taken from the closest template in the bank.
We record the detector-frame chirp mass, component masses, and aligned spin.
For the \cwb pipeline, which does not use an explicit waveform model and, therefore, does not provide these estimates, we record only the \ac{FAR} and \ac{SNR}.
Additional features for both templated and non-templated searches also exist, and in a production implementation, a full study of their importance should be made.
Finally, we use the logarithm of the \ac{IFAR} per pipeline instead of the \ac{FAR} since the significance of changes in the \ac{FAR} matters on a logarithm rather than a linear scale.
This choice also ensures that a value of zero naturally fits with the choice of zero applied to missing data.

\subsection{Conformal prediction}
\ac{CP} uses a non-conformity measure to calculate a non-conformity score $s_i^Y = A(\vec{X}_i, Y)$ for each sample $\vec{X}_i$ and label $Y$.
During calibration, the non-conformity scores for all samples in the calibration dataset (a subset of the training data) are calculated and sorted in ascending order.
In this work, we use Mondrian or label-conditional \ac{CP} \citep{vovk2013conditional, ding:2023} and so for each label $Y$ we calculate the $1-\alpha$ quantile
\begin{equation}
    \hat{q}_Y = s^Y_{\lceil(N_c^Y+1)(1-\alpha)\rceil}\,,
    \label{eq:qhat}
\end{equation}
where $N_c^Y$ is the number of data points in the calibration data with label $Y$ and $\alpha$ is the user-chosen error rate.
In the prediction step, for a sample $\vec{X}'$, we compute the non-conformity scores for each label and compute the prediction set $\Gamma^{\alpha}=\{Y: A(\vec{X}', Y) \le \hat{q}_Y\}$ where $Y$ runs over all possible labels.
The guarantee of \ac{CP} is that, in the limit that the size of the calibration data is sufficiently large, the true label $\hat{y}$ is included in the prediction set with a probability of approximately one minus the error rate:
\begin{align}
    P(\hat{y} \in \Gamma^{\alpha}) \approx 1-\alpha\,.
\end{align}

\subsection{Additional studies}

\begin{figure}
    \centering
    \includegraphics{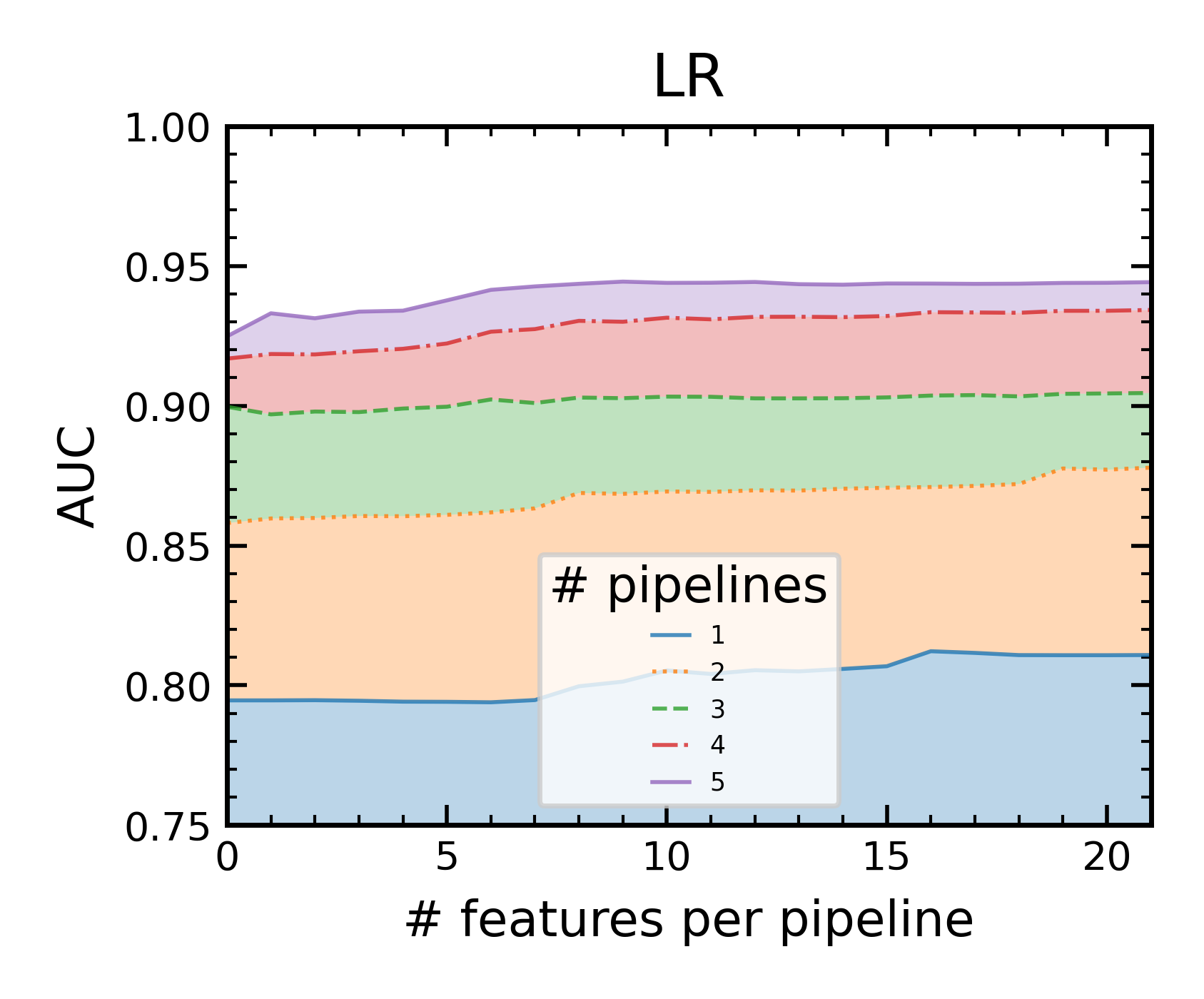}
    \includegraphics{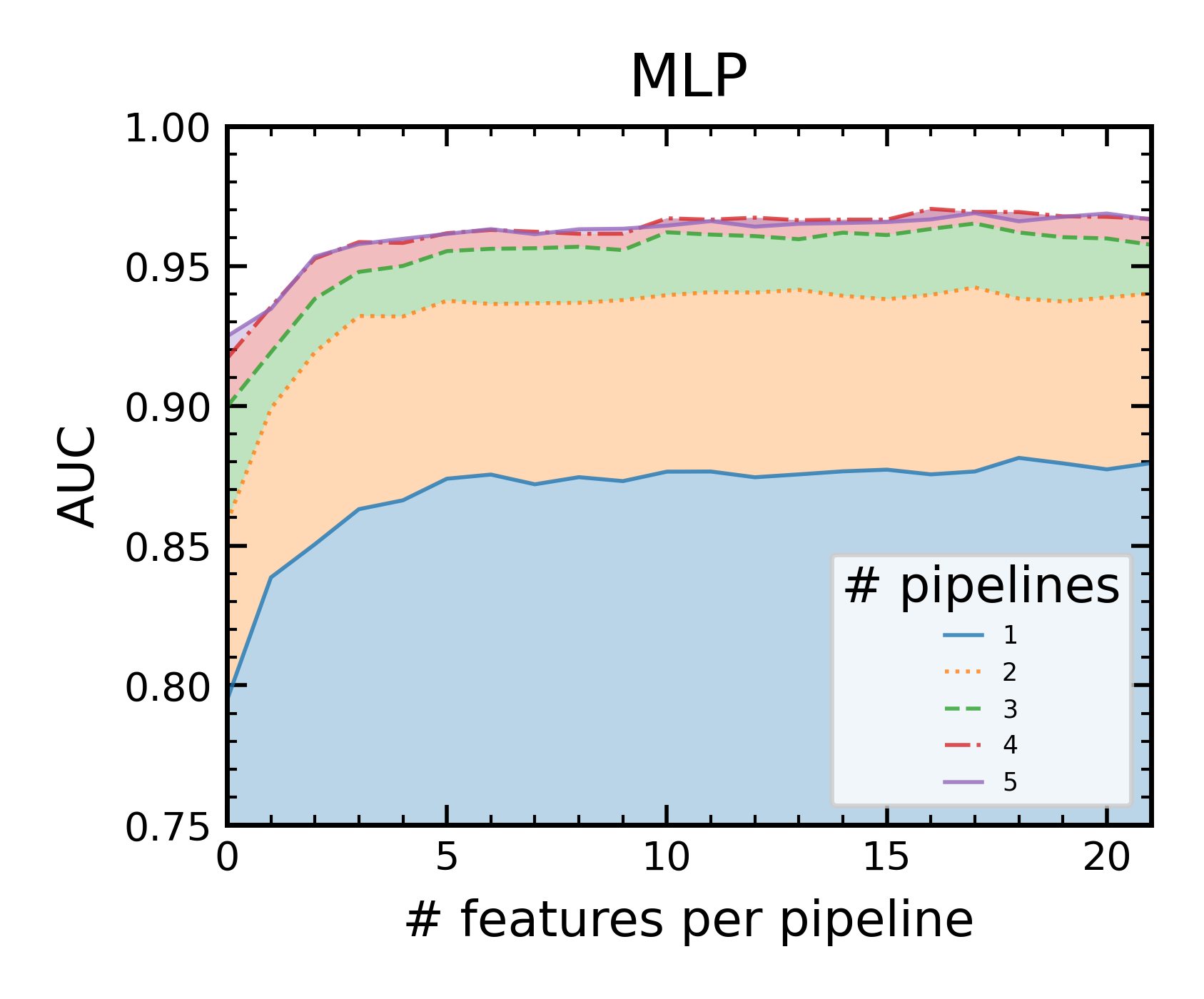}
    \caption{
    The 90\% upper limit on the measured \ac{AUC} for the \ac{LR} and \ac{MLP} calculated on different permutations of the data split, the pipelines included in the feature set, and the per-pipeline feature set.
    We show curves averaged over the number of pipelines and as a function of the number of parameters (with an ordering that reflects a choice of the likely importance, starting with the \ac{IFAR}, mass of the binary, \ac{SNR}, etc.).
    We note, however, that \cwb does not produce estimates of all features; in this case, empty rows are provided, adding no additional information.
    }
    \label{fig:auc}
\end{figure}

\emph{Feature importance} --- To further explore the importance of the number of features, in \cref{fig:auc}, we repeat the \ac{ROC} analysis of the \ac{LR} and \ac{MLP} model under different permutations of the test data and vary the number of pipelines and pipeline features included in the feature data set.
For \ac{LR} and \ac{MLP}, this figure demonstrates the importance of multiple pipelines: going from one to two pipelines produces a $\approx 5\%$ relative increase in the measured AUC in both cases, and adding more pipelines produces more moderate but still non-negligible increases in the AUC.
Meanwhile, for \ac{LR} we find negligible changes in the \ac{AUC} as more features are added to the data.
But for \ac{MLP}, \cref{fig:auc} demonstrates that, of the 21 per-pipeline features we include, only about 5 are required.
This can be understood because many of the features (e.g. the \ac{SNR} and \ac{FAR}) are highly correlated.
A detailed study of the feature set could be done to identify a limited number of the most important features, which could aid in interpreting results.

\emph{\ac{LR} interpretability} -- To demonstrate the ease of interpretation for the \ac{LR} model trained on the \ac{MDC} data, in \cref{tab:logistic_coef}, we list the top-five features 
as ranked by their weights from the \ac{LR} combination algorithm.
This demonstrates that the \ac{IFAR} and \ac{SNR} are highly-ranked features and that contributions from all pipelines are of importance.
We also find that features such as the $\chi^2$ veto statistic play an important role and have large negative coefficients (of order $\sim-1$ for most pipelines). 
However, we caution that it is the absolute value of the weights that should be used to define importance, since the signs can often be unintuitive.

\begin{table}
    \centering
    \begin{tabular}{l|l|c}
         Pipeline & Feature & Weights \\ \hline
         \gstlal & $\log_{10}(\textrm{IFAR})$ & 4.3 \\
         \pycbc & $\log_{10}(\textrm{IFAR})$  & 2.8 \\
         \cwb & $\log_{10}(\textrm{IFAR})$  &  2.5 \\
         \mbta & $\log_{10}(\textrm{IFAR})$ & 2.4 \\
         \spiir & $\textrm{SNR}_{\rm L}$ & 2.1 \\
    \end{tabular}
    \caption{The top-five fitted coefficients from the \ac{LR} model applied to the test data in Figure 1 of the main manuscript. Note that $\textrm{SNR}_{\rm L}$ refers to the measured \ac{SNR} in the Livingston detector.
    We caution that this data should not be taken as a representative ranking of the pipelines themselves since they were still under development during the analysis of the \ac{MDC} data, the numeric values depend on internal normalisations, and the importance of different features will depend on the details of the combination algorithm.
    }
    \label{tab:logistic_coef}
\end{table}

\begin{figure}
    \centering
    \includegraphics{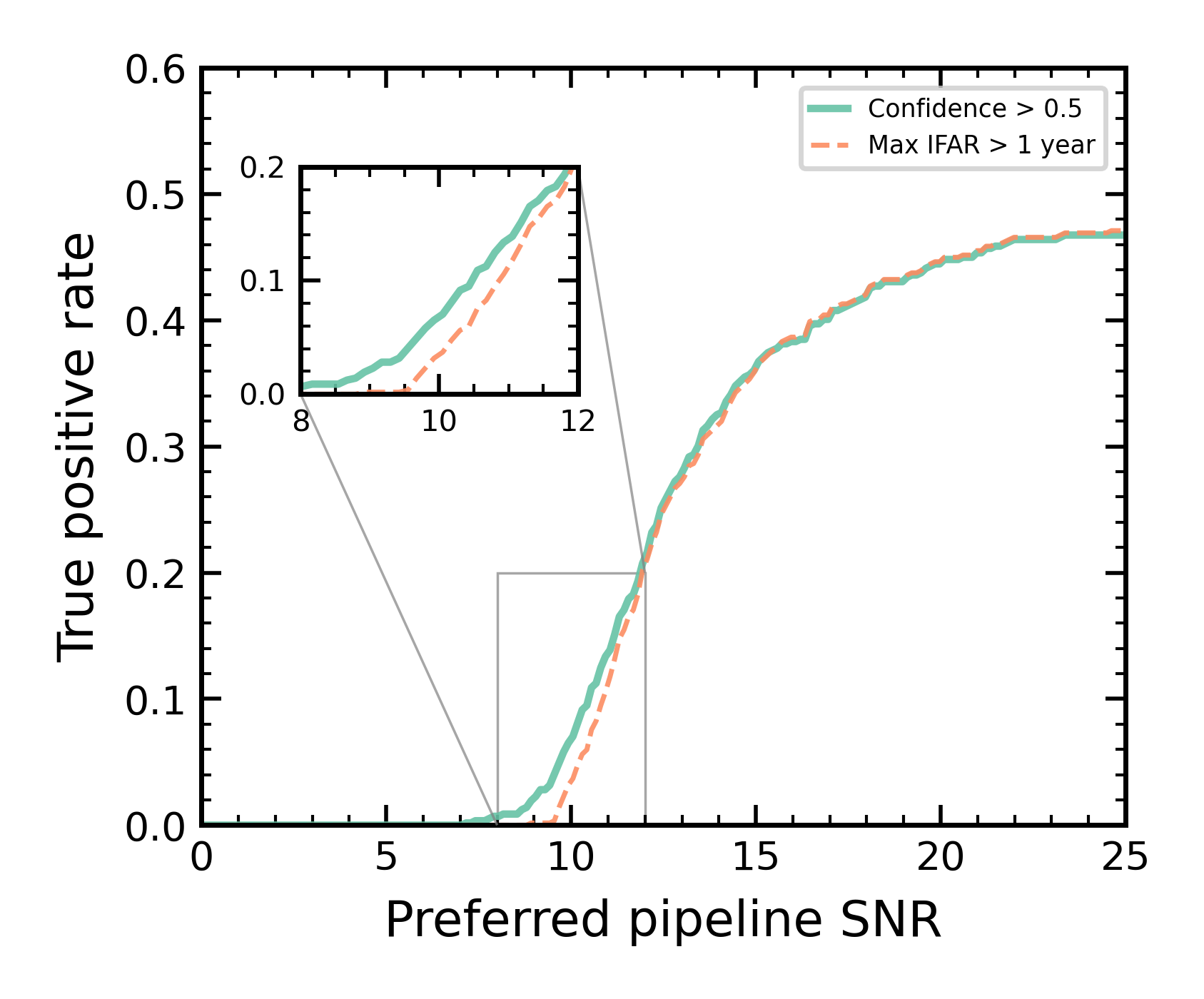}
    \caption{
    The cumulative histogram of the true positive rate against the preferred-pipeline SNR using the \ac{LR} confidence as a threshold (green solid curve) and the maximum-\ac{IFAR} (orange dashed line).
    For the \ac{LR} pipeline combination approach, we set a threshold of conditional confidence in the signal label greater than 0.5.
    For the maximum-\ac{IFAR} pipeline combination approach, we set a threshold of 1~year.
    These thresholds are arbitrarily chosen and happen to approximately match at the maximum \ac{SNR}, which helps elucidate where they differ at lower \ac{SNR} (shown in the inset axis).
    }
    \label{fig:SNR}
\end{figure}

\emph{Confidence vs. SNR} --- To further compare the combined ML pipeline with the maximum IFAR approach, in \cref{fig:SNR}, we plot the cumulative true positive rate (applying a threshold on the confidence or maximum IFAR) as a function of the preferred-pipeline SNR.
For the chosen thresholds, the two methods achieve a similar true positive rate up to the maximum observed SNR (this in itself is not meaningful, but does provide some insight into the relationship between the confidence of FAR).
However, what \cref{fig:SNR} demonstrates is that the confidence-based approach outperforms the maximum-IFAR method for SNRs around 10 (as highlighted in the inset axis), demonstrating that gains arise in building confidence in weak signals due to the coherent detection across pipelines.

\emph{Table of O3a and O3b candidates from GWTC-3} -- 
In \cref{tab:gwtc}, we tabulate the top-50 ranked candidates from O3a and O3b studied in this work, enabling a comparison of the \pastro and conditional confidence.
\begin{table*}[h]
    \centering
    \begin{tabular}{l|l|l|c|c|c}
    Candidate & Data & Pipeline & $\mathcal{M}$ & \pastro & Confidence \\ \hline
    GW200311\_115853 & O3a & \gstlal & 32.9 & 1.00 & 0.99\\
GW200115\_042309 & O3a & \mbta & 2.6 & 1.00 & 0.91\\
GW191129\_134029 & O3a & \gstlal & 8.5 & 1.00 & 0.97\\
GW190828\_063405 & O3b & \gstlal & 31.1 & 1.00 & 0.95\\
GW190728\_064510 & O3b & \gstlal & 10.5 & 1.00 & 0.86\\
GW190408\_181802 & O3b & \gstlal & 23.4 & 1.00 & 0.91\\
GW190707\_093326 & O3b & \gstlal & 9.8 & 1.00 & 0.87\\
GW191222\_033537 & O3a & \gstlal & 29.2 & 1.00 & 0.97\\
GW190814\_211039 & O3b & \gstlal & 6.4 & 1.00 & 0.73\\
GW190727\_060333 & O3b & \gstlal & 38.7 & 1.00 & 0.87\\
GW190924\_021846 & O3b & \gstlal & 6.5 & 1.00 & 0.79\\
GW190720\_000836 & O3b & \gstlal & 10.4 & 1.00 & 0.77\\
GW200112\_155838 & O3a & \gstlal & 28.7 & 1.00 & 0.34\\
GW190915\_235702 & O3b & \gstlal & 31.0 & 1.00 & 0.87\\
GW190828\_065509 & O3b & \pycbc & 17.0 & 1.00 & 0.73\\
GW190708\_232457 & O3b & \gstlal & 15.4 & 1.00 & 0.28\\
GW190930\_133541 & O3b & \pycbc & 10.2 & 1.00 & 0.58\\
GW190910\_112807 & O3b & \gstlal & 38.7 & 1.00 & 0.27\\
GW190620\_030421 & O3b & \gstlal & 24.0 & 0.99 & 0.25\\
GW190803\_022701 & O3b & \gstlal & 27.1 & 0.93 & 0.48\\
GW200302\_015811 & O3a & \gstlal & 33.4 & 0.91 & 0.23\\
GW191219\_163120 & O3a & \pycbc & 4.7 & 0.82 & 0.32\\
GW191103\_012549 & O3a & \pycbc & 10.1 & 0.77 & 0.48\\
GW190917\_114630 & O3b & \gstlal & 4.2 & 0.74 & 0.16\\
GW190425\_081805 & O3b & \gstlal & 1.5 & 0.69 & 0.22\\
GW200218\_100521 & O3a & \cwb & - & 0.49 & 0.35\\
GW200219\_201407 & O3a & \mbta & 6.1 & 0.48 & 0.31\\
GW190529\_122222 & O3b & \pycbc & 9.3 & 0.43 & 0.29\\
GW191126\_115259 & O3a & \pycbc & 11.4 & 0.39 & 0.58\\
GW191210\_141621 & O3a & \mbta & 8.1 & 0.39 & 0.16\\
GW200105\_162426 & O3a & \gstlal & 3.6 & 0.36 & 0.19\\
GW190711\_030756 & O3b & \gstlal & 29.2 & 0.35 & 0.16\\
GW190612\_115526 & O3b & \pycbc & 6.5 & 0.34 & 0.27\\
GW191107\_042137 & O3a & \mbta & 6.7 & 0.32 & 0.18\\
GW200221\_142805 & O3a & \gstlal & 40.6 & 0.29 & 0.20\\
GW191222\_135709 & O3a & \pycbc & 6.6 & 0.29 & 0.41\\
GW190810\_134240 & O3b & \gstlal & 5.1 & 0.28 & 0.11\\
GW200202\_072036 & O3a & \pycbc & 4.5 & 0.28 & 0.43\\
GW190531\_023648 & O3b & \gstlal & 2.0 & 0.28 & 0.43\\
GW190919\_144431 & O3b & \gstlal & 3.7 & 0.25 & 0.11\\
GW190907\_212802 & O3b & \gstlal & 12.3 & 0.24 & 0.11\\
GW191229\_124436 & O3a & \gstlal & 6.7 & 0.24 & 0.11\\
GW200308\_173609 & O3a & \mbta & 45.2 & 0.24 & 0.46\\
GW200223\_180658 & O3a & \gstlal & 24.1 & 0.23 & 0.16\\
GW200111\_174754 & O3a & \pycbc & 6.7 & 0.22 & 0.31\\
GW190809\_200545 & O3b & \pycbc & 5.1 & 0.20 & 0.26\\
GW200214\_031723 & O3a & \pycbc & 7.5 & 0.20 & 0.27\\
GW190625\_010322 & O3b & \pycbc & 13.3 & 0.20 & 0.26\\
GW200311\_103121 & O3a & \pycbc & 1.2 & 0.19 & 0.66\\
GW191108\_145308 & O3a & \pycbc & 6.5 & 0.17 & 0.26

    \end{tabular}
    \caption{A table of the top 50 candidates ranked by the maximum \pastro taken from the O3a and O3b candidates released in GWTC-3. The reported chirp mass is taken from the pipeline with the maximum \pastro value. For \cwb, the chirp mass is not reported, as we do not use this in our analysis.}
    \label{tab:gwtc}
\end{table*}

\subsection{Suggested improvements}
Many significant improvements could be made in this approach.
First, there are several caveats regarding the use of the MDC for training and calibration data: the rate of events is far greater than astrophysically expected, the pipelines were still under development, and the underlying data potentially includes undetected signals that could contaminate the training data.
Therefore, a more realistic MDC is needed to address these issues and provide a data set where exchangeability with real data is better assured.
Moreover, the minimum \ac{IFAR} of triggers in our test data was limited by the specification of the MDC data study, which results in just 66 multi-pipeline noise triggers in our test data (an issue also faced by \citet{Banagiri:2023ztt}).
Therefore, this should add caution to the interpretation of our results.
Hence, we recommend that when running a more realistic MDC, we also record and store the results of multiple pipelines for triggers down to a lower \ac{IFAR} threshold.
Second, we utilised a limited feature set in this work.
We would like to include intermediate data products in future work and extend the feature set to include the Bayesian \pastro and its constituent elements.
Third, we applied a simple binary classification (signal or noise), but the framework naturally extends to multi-class classification. We would like to include the source type (e.g., BBH, BNS, NSBH) and incorporate information about detector performance using detector characterisation tools.
Fourth, within the framework, additional information and scientific expertise can be incorporated into the model, e.g., by weighting the pipeline outputs during training.
Finally, in this work, we explore two \ac{ML} approaches (\ac{LR} and \ac{MLP}). Overall, we prefer the former approach because it provides an easy mechanism for understanding the importance of individual features to the final classification.
However, more involved \ac{NN} approaches offer the capacity to learn more complex correlations between features, thereby enhancing the potential for greater detection efficiency.
If we are to be successful in using these in production, though, we need to ensure that the results remain interpretable and that we understand why a given confidence is calculated in detail.

\end{document}